\documentclass[12pt,preprint]{aastex}

\newcommand{\dtau}[1][]{\frac{\textrm{d}#1}{\textrm{d}\tau}}
\newcommand{\bdtau}[1][]{ \left ( \dtau[#1] \right ) }
\newcommand{\scale}{ a(\tau) }
\newcommand{\ro}[2][0]{\ensuremath{ \rho^{#1}_{#2}}}
\newcommand{\lsim}{\lower.7ex\hbox{$\;\stackrel{\textstyle<}{\sim}\;$}}
\newcommand{\Mp}{\ensuremath{M_{\bar P}}}
\newcommand{\weff}{\ensuremath{\overline{w}_0}}
\newcommand{\ome}[2][]{ \ensuremath{
    \Omega_{#1}^{#2}}} 
\newcommand{\omebar}[2][]{\ensuremath{
    \overline{\Omega}_{#1}^{#2}}}

\shorttitle{Quintessence and CMB Peaks}
\shortauthors{Doran et al.}

\begin{document}

\title{Quintessence and the Separation of CMB Peaks}

\author{Michael Doran, Matthew Lilley, Jan Schwindt and Christof
  Wetterich} \affil{Institut f\"ur Theoretische Physik der
  Universit\"at Heidelberg, Philosophenweg 16, D-69120 Heidelberg,
  Germany}

\email{M.Doran@thphys.uni-heidelberg.de}
\email{M.Lilley@thphys.uni-heidelberg.de}
\email{J.Schwindt@thphys.uni-heidelberg.de}
\email{C.Wetterich@thphys.uni-heidelberg.de}

\begin{abstract}
  We propose that it should be possible to use the CMB to discriminate
  between dark energy models with different equations of state,
  including distinguishing a cosmological constant from many models of
  quintessence.  The separation of peaks in the CMB anisotropies can
  be parametrised by three quantities: the amount of quintessence
  today, the amount at last scattering, and the averaged equation of
  state of quintessence.  In particular, we show that the CMB peaks
  can be used to measure the amount of dark energy present before last
  scattering.
\end{abstract}

\keywords{cosmic microwave background---cosmology: theory}

\section{Introduction}

The idea of quintessence was born \citep{cw1} from an attempt to
understand the vanishing of the cosmological constant.  It was
proposed that the cosmological evolution of a scalar field may
naturally lead to an observable, homogeneous dark energy component
today.  This contrasts with the extreme fine-tuning needed in order
for a cosmological constant to become significant just at recent
times.  If quintessence constitutes a major part of the energy density
of the Universe today, say $\ome[0]{\phi} > 0.5$, structure formation
tells us that this cannot always have been so in the past
\citep{peeb88,peeb88_2,fj,fj2}.  Combining the phenomenology of a
large quintessence component with the quest for naturalness
\citep{he00} leads to cosmologies with an equation of state for
quintessence changing in time, compatible with a universe accelerating
today. In several aspects of phenomenology the models with a dynamical
dark matter component resemble the cosmology with a cosmological
constant \citep{steinhardt}.  It is therefore crucial to find possible
observations which allow us to discriminate between the dynamical
quintessence models and a constant dark energy theory (i.e. a
cosmological constant).  The detailed structure of the anisotropies in
the cosmic microwave background radiation (CMB) depends upon two
epochs in cosmology: around the emission of the radiation (last
scattering) and today.  The CMB may therefore serve as a test to
distinguish models where quintessence played a role at the time of
last scattering from those where it was insignificant at this epoch.
It may also reveal details of the equation of state of quintessence
(characterised by $w = p^{\phi} / \rho^{\phi}$) in the present epoch.

The calculation of CMB spectra is, in general, an elaborate task
\citep{se96,hu95}. However, the location of the peaks and, for our
purpose, the \emph{spacing} between the peaks can be estimated with
much less detailed knowledge if adiabatic initial conditions and a
flat universe are assumed. The oscillations of the primeval plasma
before decoupling lead to pronounced peaks in the dependence of the
averaged anisotropies on the length scale. When projected onto the sky
today, the spacing between the peaks at different angular momentum $l$
depends, in addition, on the geometry of the universe at later time.
It is given, to a good approximation, by the simple formula
\citep{hu95,hu97}
\begin{equation}
\label{platz}
\Delta l = \pi  \frac{\tau_0 - \tau_{\rm ls}}{s} = \pi \frac{\tau_0 -
  \tau_{\rm ls}}{\bar c_s \tau_{\rm ls}}.
\end{equation}
Here $\tau_0$ and $\tau_{\rm ls}$ are the conformal time today and at
last scattering (which are equal to the particle horizons) and
$\tau=\int {\rm d}t \ a^{-1}(t)$, with cosmological scale factor $a$.
The sound horizon at last scattering $s$ is related to $\tau_{\rm ls}$
by $s =\bar c_s\tau_{\rm ls}$, where the average sound speed before
last scattering $\bar c_s \equiv \tau_{\rm ls}^{-1} \int_0^{\tau_{\rm
    ls}} {\rm d}\tau\ c_s$ obeys $c_s^{-2} = 3 + (9/4)(\rho^{\rm
  b}(t)/\rho^{\rm r}(t))$, with $\rho^{\rm b}/\rho^{\rm r}$ the ratio
of baryon to photon energy density. We note a direct dependence of 
$\Delta l$ on the present geometry through $\tau_0$ as well as an indirect
one through the dependence of $\tau_{ls}$ on the amount of dark energy
today (see Equation (\ref{tauls})).

The {\em location} of the $m-$th peak can be approximated by
\citep{hfzt}
\begin{equation}\label{platz_2}
l_m = \Delta l \left(m - \phi\right),
\end{equation}
where the phase-shift $\phi$ is typically less than $0.4$ and is
determined predominantly by recombination physics. By taking the ratio
of two peak locations (say $l_1/l_2$), the factor $\Delta l$ and with
it the dependence on post-recombination physics drops out and we are
in principle able to probe pre-recombination dark energy directly.  If
the other cosmological parameters were known the dependence of
$\phi$ on the amount of dark energy at last scattering could provide a
direct test of this aspect of quintessence models.  Unfortunately,
$\phi$ also depends on other cosmological parameters including the
baryon density and spectral index and there is no known analytic
formula for $\phi$ (and in fact $\phi$ does have some $m-$dependence).
We first concentrate on the peak spacing $\Delta l$ for which an
analytic formula can be given.

The equation of state of a hypothetical dark energy component
influences the expansion rate of the Universe and thus the locations
of the CMB peaks \citep{fj,fj2,coble,caldwell,luca}. In particular the
horizons at last scattering and today are modified, leaving an imprint
in the spacing of the peaks.  The influence of dark energy on the
present horizon and therefore on the CMB has been discussed in
\citep{brax00}.  A likelihood analysis on combined CMB, large scale
structure and supernovae data \citep{efstat,bond00} can also give
limits on the equation of state.
Several of these analysis concentrate on models where the
dark energy component is negligible at last scattering. In contrast,
we are interested particularly in getting information about
dark energy in early cosmology. Therefore, the amount of dark energy
at last scattering is an important parameter in our investigation.

\begin{figure}[!ht]
\begin{center}
\input{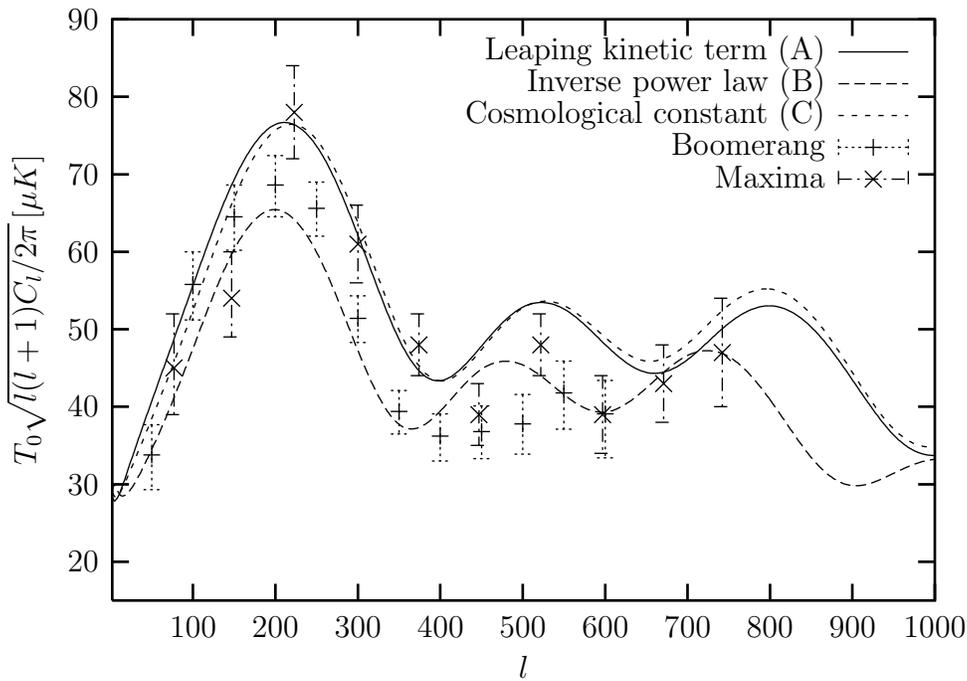}
\end{center}
\caption{The CMB Spectrum for $\Lambda$-CDM (model C), leaping kinetic term
  (model A) and inverse power law (model B) quintessence universes
  with $\ome[0]{\phi}=0.6$.  The data points from the Boomerang
  \citep{boom} and Maxima \citep{maxima} experiments are shown for
  reference. \label{cmb}}
\end{figure}

We present here a quantitative discussion of the mechanisms which
determine the spreading of the peaks.  A simple analytic formula
permits us to relate $\Delta l$ directly to three characteristic
quantities for the history of quintessence, namely the fraction of
dark energy today, $\ome[0]{\phi}$, the averaged ratio between dark
pressure and dark energy, $\weff = \langle p^{\phi} /
\rho^{\phi}\rangle_0$, and the averaged quintessence fraction before
last scattering, $\omebar[\rm ls]{\phi}$ (for details of the averaging
see below).  We compare our estimate with an explicit numerical
solution of the relevant cosmological equations using CMB-FAST
\citep{se96}.  For a given model of quintessence the computation of
the relevant parameters $\ome[0]{\phi}, \weff$ and $\omebar[\rm
ls]{\phi}$ requires the solution of the background equations.  Our
main conclusion is that future high-precision measurements of the
location of the CMB-peaks can discriminate between different models of
dark energy if some of the cosmological parameters are fixed by
independent observations.  It should be noted here that a likelihood
analysis of the kind performed in \citep{bond00}, where $w$ is assumed
to be constant throughout the history of the Universe, would not be
able to extract this information as it does not allow
$\omebar[ls]{\phi}$ to vary.  We point out that for time-varying $w$
there is no direct connection between the parameters $\weff$ and
$\omebar[ls]{\phi}$, i.e. a substantial $\omebar[ls]{\phi}$ (say 0.1)
can coexist with rather large negative $\weff$.  We perform therefore
a three parameter analysis of quintessence models and our work goes
beyond the investigation for constant $w$ in \citep{steinhardt}.

\section{CMB Peaks in Quintessence Models}

We wish first to illustrate the impact of different dark energy models
on the fluctuation spectrum of the CMB by comparing three examples.
The first corresponds to a `leaping kinetic term quintessence'
\citep{he00} (A), the second to `inverse power-law quintessence'
\citep{peeb88,peeb88_2} (B) and the third to a cosmological constant
(C).  The three examples, whose parameters are chosen such that
$\ome[0]{\phi} = 0.6$ for each, give similar predictions for many
aspects of cosmological observation (we assume everywhere a flat
universe $\Omega_{\rm total} = 1$).  Details of the models can be
found below in Section \ref{sect:examples}.  We solve the cosmology
using CMB-FAST for a flat initial spectrum with parameters specified
in Table \ref{symb}.  Figure \ref{cmb} clearly demonstrates that the
fluctuation spectra of the three models are distinguishable by future
high-precision measurements.  This can be traced back to different
values of $\omebar[\rm ls]{\phi}$ and $\weff$, namely $\omebar[\rm
ls]{\phi} = (0.13,0,0)$ and $\weff = (-0.45, -0.37, -1)$ for models
(A,B,C).  These quantities enter a simple analytic formula (derived
below in Section \ref{sect:estimate}) for the spacing between the
peaks
\begin{equation}
\label{sep}
\Delta l =   \pi \bar c_s^{-1} \left[
      \frac{F(\ome[0]{\phi},\weff)}{\sqrt{1-{\omebar[\rm ls]{\phi}}}}
      \left \{ \sqrt{a_{\rm ls} +
      {\ome[0]{\rm r}\over 1 - \ome[0]{\phi}}} - \sqrt{{\ome[0]{\rm r}\over
      1 - \ome[0]{\phi}}} \right \} ^{-1} - 1 \right ],
\end{equation}
with
\begin{equation} \label{F_int}
F(\ome[0]{\phi},\weff) = \frac{1}{2} \int_0^1 \textrm{d}a \left( a +
  \frac{\ome[0]{\phi}}{1-\ome[0]{\phi}} \, a ^{(1 - 3 \weff)} +
  \frac{\ome[0]{\rm r}(1-a)}{1-\ome[0]{\phi}} \right)^{-1/2},
\end{equation}
and today's radiation component $\ome[0]{\rm r}= 9.89\times 10^{-5},
a_{\rm ls}^{-1} = 1100$ and $\overline{c}_s = 0.52$.  In Table 1 we
evaluate Equation (\ref{sep}) for quintessence models with various
parameters, together with the locations $l_1, l_2$ of the first two
peaks computed by CMB-FAST.  The last entry contains the peak spacing
averaged over 6 peaks for the numerical solution.  This demonstrates
that an accurate measurement of the peak spacing $\Delta l$ is a
powerful tool for the discrimination between different dark matter
models!

\section{Analytic Estimate of Peak Spacing} \label{sect:estimate}
We derive next the formula (\ref{sep}).  Our first task is to estimate
the sound horizon at decoupling.  We assume that the fraction of
quintessential energy $\ome[]{\phi}(\tau)$ does not change rapidly
for a considerable period before decoupling and define an effective
average ${\omebar[\rm ls]{\phi}} \equiv \tau_{\rm ls}^{-1}
\int_0^{\tau_{\rm ls}} \ome[]{\phi}(\tau) \textrm{d}\tau$.  We note
that this average is dominated for $\tau$ near $\tau_{\rm ls}$ whereas
very early cosmology is irrelevant.  Approximating $\ome[]{\phi}$ by
the constant average ${\omebar[\rm ls]{\phi}}$ for the period around
last scattering, the Friedmann equation for a flat universe reads
\begin{equation}
\label{friedmann_ls}
3 \Mp^2 H^2(t) (1- {\omebar[\rm ls]{\phi}}) = 
\ro[\rm m]{}(t) + \ro[\rm r]{}(t) =
\ro[\rm m]{0} a(t)^{-3} + \ro[\rm r]{0} a(t)^{-4}.
\end{equation}
Here $\Mp = (8 \pi G_N)^{(-1/2)}$ is the reduced Planck mass, $H(t)$
is the Hubble parameter and \ro[\rm m]{0} and \ro[\rm r]{0} are the matter and
relativistic (photons and 3 species of neutrinos) energy densities
today.

Today, neglecting radiation, we have $3 \Mp^2 H_0^2 (1 -
\ome[0]{\phi}) = \ro[\rm m]{0}$, which we insert in Equation
(\ref{friedmann_ls}) to obtain
\begin{equation}
\bdtau[a]^2 = H_0^2 (1-{\omebar[\rm ls]{\phi}})^{-1} \left [ (1-
  \ome[0]{\phi}) \scale + \ome[0]{\rm r} \right ],
\end{equation}
where we have changed from coordinate time $t$ to conformal time
$\tau$.  Separating the variables and integrating gives
\begin{equation}
\label{tauls}
\tau_{\rm ls} = 2 H_0^{-1} \sqrt{
  \frac{1-{\omebar[\rm ls]{\phi}}}{1-\ome[0]{\phi}}} \left \{
  \sqrt{a_{\rm ls} 
    + \frac{\ome[0]{\rm r}}{1 - \ome[0]{\phi}}} - \sqrt{
    \frac{\ome[0]{\rm r}}{1 - \ome[0]{\phi}}} \right \},
\end{equation}
which is well known for vanishing ${\omebar[\rm ls]{\phi}}$.  For fixed
$H_0,\ \ome[0]{\phi},\ \ome[0]{\rm r}$ and $a_{\rm ls}$ (see Table
\ref{symb} for the values used in this paper), we see that $\tau_{\rm
  ls} = \tau_{\rm ls}^{\rm vac} \, (1
-{\omebar[\rm ls]{\phi}})^{(1/2)}$, where $\tau_{\rm ls}^{\rm vac}$ is
the last scattering horizon for a \mbox{$\Lambda$-CDM} universe (which
we treat here to be just a special realisation of dark energy with
$w=-1$).  To estimate the sound horizon, we also need $\bar c_s$,
which may be obtained numerically and in our model universe is $0.52$.

Turning to the horizon today, we mimic the steps of above, this time
assuming some equation of state $p^{\phi}(t) = w(t) \rho^{\phi}(t)$
for quintessence.

\begin{figure}[!ht]
\begin{center}
\input{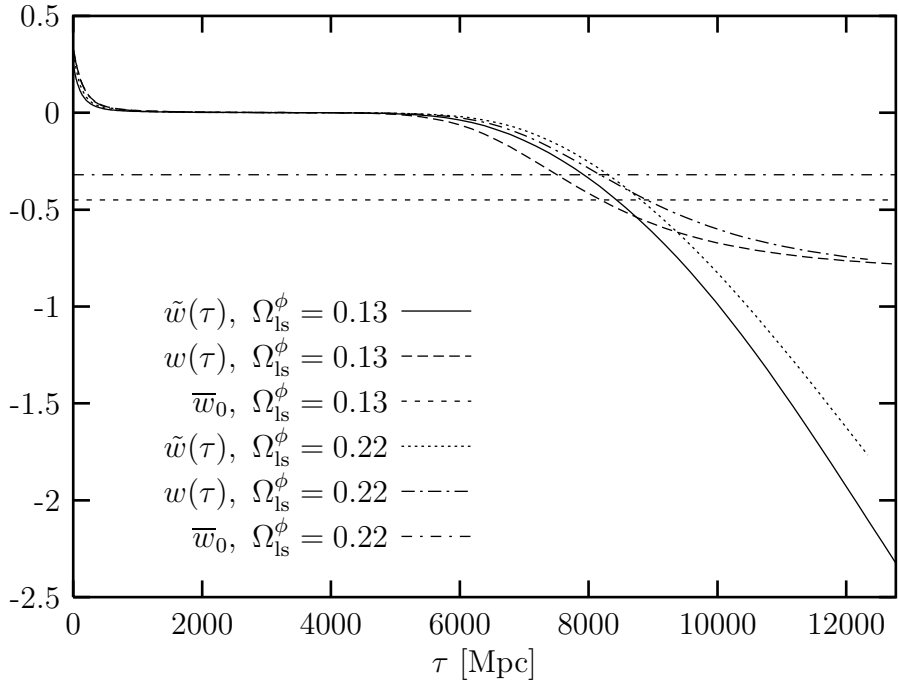}
\end{center}
\caption{Equation of state $w(\tau)$, $\tilde{w}(\tau) \equiv
  \ome[]{\phi}(\tau) w(\tau) \tau_0/ \int_0^{\tau_0}
  \ome[]{\phi}(\tau') \textrm{d} \tau' $ and averaged equation of
  state $\weff$ for the leaping kinetic term model with
  $\omebar[\rm ls]{\phi} = 0.13$ and $\omebar[\rm ls]{\phi} = 0.22$. \label{effectiveW}}
\end{figure}

We define an averaged value $\weff$ by  
\begin{equation}
\label{w_eff}
\weff =  \int_0^{\tau_0} \ome[]{\phi}(\tau) w(\tau) \textrm{d} \tau 
\times \left(  \int_0^{\tau_0} \ome[]{\phi}(\tau) \textrm{d} \tau
\right)^{-1}.
\end{equation}
It is $\ome[]{\phi}$-weighted, reflecting the fact that the equation
of state of the dark energy component is more significant if the dark
energy constitutes a higher proportion of the total energy of the
Universe (see Figure \ref{effectiveW}).

In the limiting case that the equation of state did not change during
the recent history of the Universe, the average is of course equal to
$w$ today.  Nevertheless, the difference between the average $\weff$
and today's value $w_0$ can be substantial for certain models, as can
be seen from Table \ref{horizonte}.

Integrating the cosmological equation with constant $\weff$
\begin{equation}
\bdtau[a]^2 = H_0^2  \left \{   (1- \ome[0]{\phi}- \ome[0]{\rm r})
  \scale +  \ome[0]{\phi}
  a^{(1-3 \weff)} + \ome[0]{\rm r}  \right \},
\end{equation}
gives
\begin{equation}
\label{tau0}
\tau_0 = 2 H_0^{-1}    (1- \ome[0]{\phi})^{-\frac{1}{2}}
F(\ome[0]{\phi},\weff), 
\end{equation}
with $F$ given by Equation (\ref{F_int}). Substituting Equations
(\ref{tauls}) and (\ref{tau0}) into Equation (\ref{platz}), we obtain
the final result (\ref{sep}).

\begin{figure}[!ht]
\begin{center}
\input{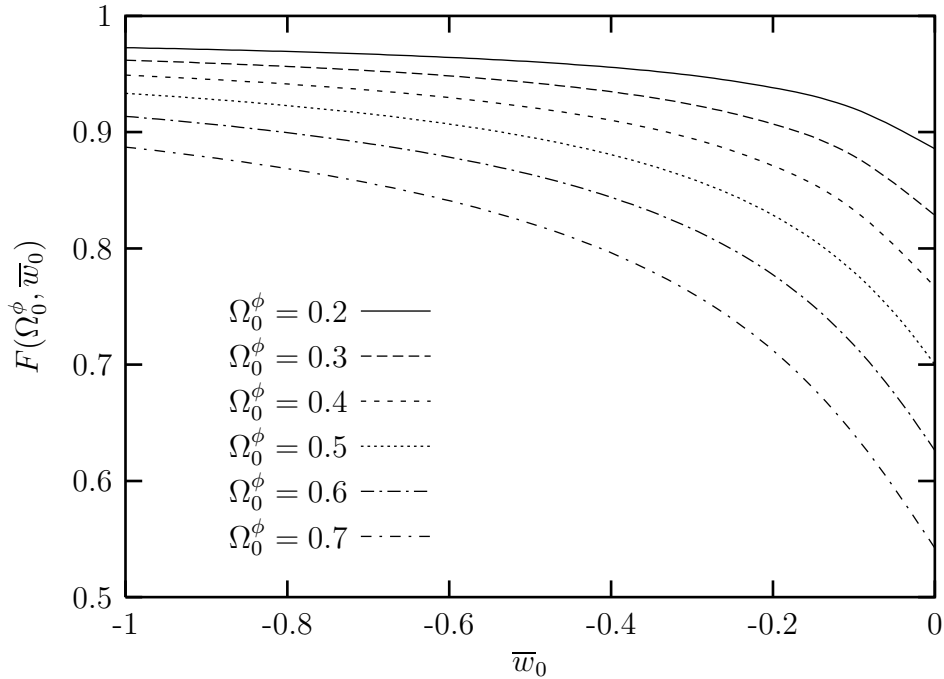}
\end{center}
\caption{$F(\ome[0]{\phi},\weff)$ as a function of $\weff$ of
  the dark energy component, for $\ome[0]{\phi}$ between 0.2 and
  0.7. Between the limiting cases of $\weff=-1$ (cosmological
  constant) and $\weff=0$ (corresponding to pressureless dust), the
  age of the Universe varies considerably. \label{f60}}
\end{figure}

The integral $F$ of Equation (\ref{F_int}) can be solved analytically
for special values of $\weff$, e.g.
\begin{equation}
F(\ome[0]{\phi},\weff=0) = \sqrt{1 - \ome[0]{\phi}}\left( 1
     - \sqrt{\ome[0]{\rm r}}\right) + O(\ome[0]{\rm r}).
\end{equation}
Since the integral (\ref{F_int}) is dominated by $a$ close to one
(typically $\weff \leq 0$) only the present epoch matters, consistent
with the averaging procedure (\ref{w_eff}). From this we regain on
inserting in Equation (\ref{tau0}) the trivial result that the age of
the Universe is the same for a cold dark matter and a pressureless
dark energy universe.  We plot $F(\ome[0]{\phi},\weff)$ for various
values of $\ome[0]{\phi}$ in \mbox{Figure \ref{f60}.}

For $\ome[0]{\phi} \lsim 0.6$, Equation (\ref{sep}) to good
approximation (better than one percent) can be written
\begin{equation}
\Delta l =   \pi \bar c_s^{-1} \left [ \frac{F(\ome[0]{\phi},\weff
    )}{\sqrt{ a_{\rm ls}(1-{\omebar[\rm ls]{\phi}} ) }}  \left \{ 1 +
    \left (
    \frac{\ome[0]{\rm r}}{a_{\rm ls}\left(1-\ome[0]{\phi}\right)}
    \right )^{1/2} + 
    \frac{\ome[0]{\rm r}}{2 a_{\rm ls}\left(1-\ome[0]{\phi}\right)}
    \right\}-  1 \right ].
\end{equation}

The precision of our analytic estimate for $\Delta l$ can be inferred
from Table \ref{lollie}. Similarly, we show in Table \ref{horizonte}
the accuracy of the estimates of $\tau_{\rm ls}$ (\ref{tauls}) and
$\tau_0$ (\ref{tau0}) by comparison with the numerical solution. This
demonstrates that our averaging prescriptions are indeed meaningful.
We conclude that the influence of a wide class of different
quintessence models (beyond the ones discussed here explicitly) on the
spreading of the CMB-peaks can be characterised by the three
quantities $\ome[0]{\phi},\ {\omebar[\rm ls]{\phi}}$ and $\weff$.

\section{Ratios of peak locations}
An alternative to the spacing between the peaks is the ratio of any
two peak (or indeed trough) locations.  After last scattering the CMB
anisotropies simply scale according to the geometry of the Universe --
taking the ratio of two peak locations factors out this scaling and
leaves a quantity which is sensitive only to pre-last-scattering
physics.  As can be seen in Table 1, (spatially-flat) models with
negligible $\omebar[ls]{\phi}$ all have $l_2 / l_1 \approx 2.41$ for
the parameters given in Table 3.  The dependence of this ratio on the
other cosmological parameters can be computed numerically
\citep{doran}.  If the other parameters can be fixed by independent
observations, the ratios of peak locations are fixed uniquely for
models with vanishing \omebar[ls]{\phi}. A deviation from the
predicted value would be a hint of time-varying quintessence.  It may
also be possible to make a direct measurement of $\omebar[ls]{\phi}$
from ratios of successive peak locations.

\section{Specific Quintessence Models} \label{sect:examples}
Different models of quintessence may be characterised by the potential
$V(\phi)$ and the kinetic term of the scalar `cosmon'-field $\phi$
\begin{equation}
    \mathcal{L}(\phi) = {1\over 2}\left(\partial_\mu \phi\right)^2
    k^2\left(\phi\right) + V(\phi). 
\end{equation} 
For practical purposes, a variable transformation allows us to work
either with a standard kinetic term $k(\phi)=1$ or a standard
potential, i.e. $\bar V(\phi) = \Mp^4 \exp(- \phi/\Mp)$.  A
cosmological constant corresponds to the limit $V(\phi) = \lambda,\ 
k(\phi) = 0$. It is also mimicked by $k(\phi) \to \infty$.  We
consider four types of model.

\begin{description} 
\item[A.] A `leaping kinetic term' model \citep{he00}, with
  \label{eins}
  \begin{equation}
    V(\phi)=\bar V(\phi) = \Mp^4 \exp(- \phi/\Mp)
  \end{equation}
    and kinetic term 
  \begin{equation} 
    k\left(\phi\right) = k_{\rm min} + \tanh\left[\left(\phi -
    \phi_1\right)/\Mp \right]
    + 1.
  \end{equation}
  We have taken $k_{\rm min} = 0.05,\,0.1,\,0.2$ and $0.26$ and $\phi_1$ is
  adjusted to $\approx 277$ in order to obtain $\ome[0]{\phi}=0.6$.
  The value of ${\omebar[\rm ls]{\phi}}$ is determined by
  these parameters.
  
\item[B.] An inverse power law potential \citep{peeb88,peeb88_2}, with
  $k(\phi)=1$ and \label{zwei}
 \begin{equation} 
   V(\phi)=  A \phi^{-\alpha},
 \end{equation}
 We have chosen $\alpha = 6, 22$ and $40$, and $A$ adjusted such that
 $\ome[0]{\phi}=0.6$. Once again, ${\omebar[\rm ls]{\phi}}$ follows.

\item[C.] A cosmological constant tuned such that
  $\ome[0]{\ensuremath{\phi}}\equiv\ome[0]{\ensuremath{\Lambda}}
  =0.6$. 
  
\item[D.] The original exponential potential \citep{cw1,we94}, with
  $k(\phi)=1$ and \label{drei}
  \begin{equation} 
   V(\phi) =  \Mp^4 \exp(-\sqrt{2} \alpha \phi/\Mp),
  \end{equation}
  where $\alpha = \sqrt{3/2 \ome[0]{\phi}}$.

\end{description}

For the models (A) and (D), quintessence is not negligible at last
scattering. The pure exponential potential requires $\ome[0]{\phi}
\leq 0.2$ for consistency with nucleosynthesis and structure
formation. It does not lead to a presently accelerating universe.  We
quote results for $\ome[0]{\phi}=0.6$ for comparison with other models
and in order to demonstrate that a measurement of $\Delta l$ can serve
as a constraint for this type of models, independently of other
arguments. The inverse power law models (B) are compatible with a
universe accelerating today only if ${\omebar[\rm ls]{\phi}}$ is
negligible. Again, our parameter list includes cases which are not
favoured by phenomenology.  As an illustration we quote in Table 1 the
value of $\sigma_8$, which should typically range between $0.6$ and
$1.1$ for the models considered. For example, the exponential
potential model with large ${\omebar[\rm ls]{\phi}}$ is clearly ruled
out by its tiny value of $\sigma_8$\footnote{Of course $\sigma_8$
  itself also depends on other cosmological parameters and so it alone
  cannot be used to determine ${\omebar[\rm ls]{\phi}}$.}.  The main
interest for listing also phenomenologically disfavored models arises
from the question to what extent the location of the peaks can give
independent constraints.  From the point of view of naturalness, only
the models (A) and (D) do not involve tiny parameters or small mass
scales.

\begin{figure}[!ht]
\begin{center}
\includegraphics[scale=0.73]{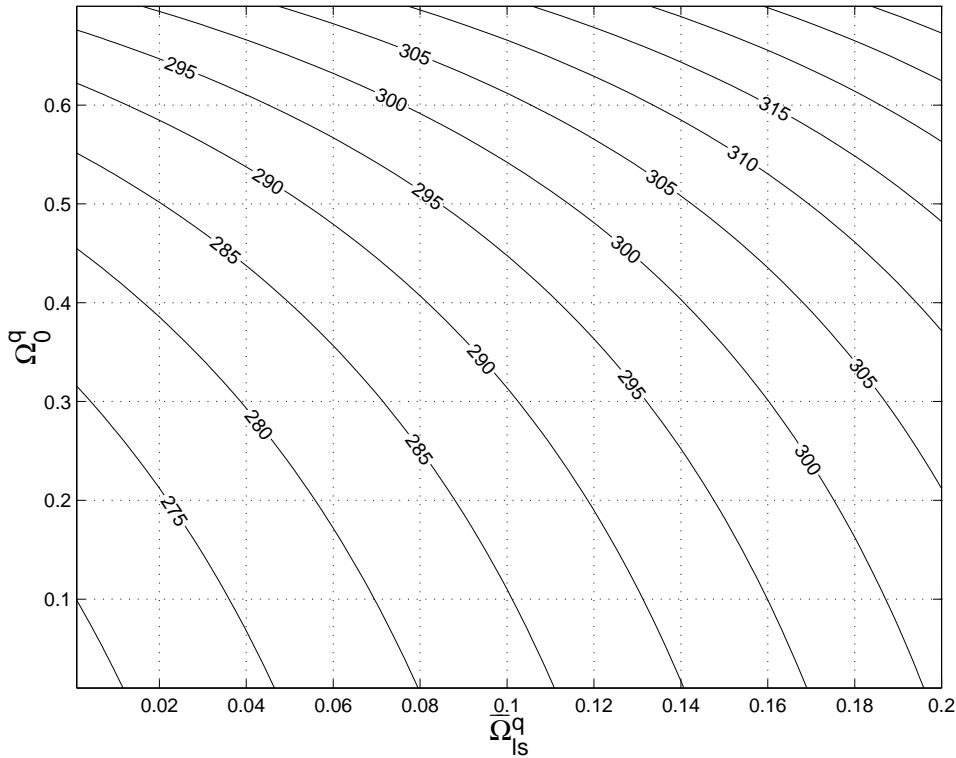}
\end{center}
\caption{Contours of equal peak spacing $\Delta l$ as a function of
  $\ome[0]{\phi}$ and ${\omebar[\rm ls]{\phi}}$.The average equation
  of state is kept fixed, $\weff = -0.7$. Increasing
  ${\omebar[\rm ls]{\phi}}$ leads to a pronounced stretching of the
  spacing. \label{mat}}
\end{figure}

The horizons and $\Delta l$ for the models considered are shown in
Tables \ref{horizonte} and \ref{lollie}.  We note that the estimate
and the exact numerical calculation are in very good agreement.  A
different choice of $a_{\rm ls}$, say $a_{\rm ls}^{-1} = 1150$, would
have affected the outcome on the low-percent level.  Also, the average
spacing obtained from CMB-FAST varies slightly (at most $2\%$) when
averaging over $4,\,5$ or $6$ peaks.  For a fixed value of the
equation of state, $\weff = -0.7$, we plot the peak spacing as a
function of $\ome[0]{\phi}$ and ${\omebar[\rm ls]{\phi}}$ in Figure
\ref{mat}.

For fixed $\weff$ and $\ome[0]{\phi}$, we see from Equation
(\ref{sep}) that $\Delta l \propto
(1-{\omebar[\rm ls]{\phi}})^{(-1/2)}$.  Hence, when combining bounds on
$\ome[0]{\phi}$ and $\weff$ from the structure of the Universe,
supernovae redshifts and other sources with CMB data, the amount of
dark energy in a redshift range of $z \sim 10^5$ to last scattering
$z\sim 1100$ may be determined.

From Figure \ref{effectiveW}, we see that the averaged equation of
state of the quintessence field for the present epoch is, in
principle, a very influential quantity in determining the spreading of
the peaks. Since combined large scale structure, supernovae and CMB
analysis in \citep{bond00} suggest $\weff \lsim -0.7$, the difference
between a cosmological constant and quintessence may be hard to spot
if ${\omebar[\rm ls]{\phi}}$ is negligible.  However, even with the data
currently available, the first peak is determined to be at $l = 212
\pm 7$ \citep{bond00}.  Once the third and fourth peak have been
measured, the measurement of the spacing between the peaks becomes an
averaging process with high precision.  We can then hope to
distinguish between different scenarios.

\section{Conclusions}
The influence of quintessence on the spacing between the CMB peaks is
determined by three quantities: $\ome[0]{\phi}$, $\omebar[\rm
ls]{\phi}$, and $\weff$.  When the location of the third peak is
accurately measured, we can hope to be able to discriminate between a
pure cosmological constant and a form of dark energy that has a
non-trivial equation of state -- possibly, and most likely, changing
in time.  The peak ratios will help determining $\omebar[\rm
ls]{\phi}$ which in principle can also be extracted from $\Delta l$,
if $\ome[0]{\phi}$ and $\weff$ are measured by independent
observations.  With $\omebar[\rm ls]{\phi}$ fixed, the peak spacing
can be used to constrain $\ome[0]{\phi}$ and $\weff$.
This can permit consistency checks for the quintessence scenario.
Together with bounds on $\ome[]{\phi}$ for the period of
structure formation ($5\lsim z \lsim 10^4$) and the bound
$\ome[BBN]{\phi} < 0.2$ from big bang nucleosynthesis
\citep{cw1,we94,ns} ($z\sim 10^9$) we will post a few milestones in
our attempt to trace the cosmological history of quintessence.

\acknowledgments

M. Doran would like to thank Luca Amendola for his kind and ongoing
support.


\clearpage



\begin{deluxetable}{cccccccc}
  \tablewidth{0pc} \tablecolumns{8} \tablecaption{Location and spacing
    of the CMB peaks for several models \tablenotemark{a}
    \label{lollie}} \tablehead{ \colhead{$ {\omebar[\rm ls]{\phi}}$} &
    \colhead{$\weff$} & \colhead{$ l_1$} & \colhead{$ l_2$} &
    \colhead{$l_2/l_1$} & \colhead{$ \Delta l ^{\rm estim.}$} 
     & \colhead{$ \Delta l^{\rm num.}$} & \colhead{$\sigma_8$}}
  \startdata 
    \cutinhead{Leaping kinetic term (A), $\ome[0]{\phi} = 0.6$} $8.4
  \times 10^{-3} $ & $-0.76$ &$215$ & $518$ & $2.41$ & $292$ &
  $291$ & $0.86$\\*
  $0.03$ & $-0.69$& $214$ & $520$ & $2.43$ & $294$ & $293$ & $0.78$\\*
  $0.13$ & $-0.45$&  $211$ & $523$ & $2.48$ &$299$  & $300$ & $0.47$ \\*
  $0.22$ & $-0.32$ & $207$ & $524$ & $2.53$ & $302$ & $307$ & $0.29$\\
  \cutinhead{Inverse power law potential (B), $\ome[0]{\phi} = 0.6$}
  $8.4 \times 10^{-8}$ & $-0.37$&$199$ & $480$ & $2.41$ & $271$ &
  $269$ & $0.61$ \\*
  $9.9 \times 10^{-2}$ & $-0.13$&$178$ & $443$ & $2.49$ & $252$ &
  $252$ & $0.18$ \\*
  $0.22$ & $-8.1\times 10^{-2}$& $172$ & $444$ & $2.58$ &$257$ &
  $257$ & $0.09$ \\
  \cutinhead{Pure exponential potential, $\ome[0]{\phi} = 0.6$} $0.70$
  & $7\times 10^{-3}$ & $190$ & $573$ & $3.02$ & $368$ &
  $377$ & $0.01$ \\
  \cutinhead{Pure exponential potential, $\ome[0]{\phi}=0.2$} $0.22$ &
  $4.7\times 10^{-3}$ & $194$ & $490$ & $2.53$ & $282$ &
  $281$ & $0.38$ \\
  \cutinhead{Cosmological constant (C), $\ome[0]{\phi} = 0.6$}
  $0$ & $-1$&  $219$ & $527$ & $2.41$ &$296$ & $295$ & $0.97$ \\
  \cutinhead{Cold Dark Matter - no dark energy, \ome[0]{\phi} = 0}
  $0$ & \nodata&  $205$ & $496$ & $2.42$ &$269$ & $268$ & $1.49$\\
  \enddata \tablenotetext{a}{The analytic estimate of $\Delta l$ stems
    from Equation (\ref{sep}). The position of the peaks $l_1$ and
    $l_2$ and $\Delta l^{\rm num.}$ are calculated with a modified
    CMB-FAST code.}
\end{deluxetable}

\begin{deluxetable}{cccccccc}
  \tablewidth{0pc} \tablecolumns{8} \tablecaption{Horizons in Mpc at
    last scattering and today for various kinds of quintessence. \tablenotemark{a} \label{horizonte}}
\tablehead{
\colhead{$ {\omebar[\rm ls]{\phi}}$} &  \colhead{$w_0$} & \colhead{\weff} &
\colhead{$\tau_0^{\rm estim.}$} & \colhead{$\tau_0^{\rm num.}$} 
& \colhead{$\Delta \tau_0$} & \colhead{$\tau_{\rm ls}^{\rm estim.}$} 
& \colhead{$\Delta \tau_{\rm ls}$} }
\startdata
\cutinhead{Leaping kinetic term (A), $\ome[0]{\phi} = 0.6$}
$8.4\times 10^{-3}$ & $-0.79$ &$-0.76$ & $ 13073$  & $13060$ & $0.1\%$
&$266$ & $0.3\%$\\*
$0.03$ & $-0.79$ & $-0.69$ & $ 12971$  & $13000$ & $0.2\%$ &$263$ &
$0.3\%$\\*
$0.13$ & $-0.78$ & $-0.45$ & $ 12470$  &  $12590$ & $1.0\%$ &$248$ &
$0.2\%$\\*
$0.22$ & $-0.75$ & $-0.32$ & $ 12012$ & $12175$ &$1.3\%$ & $236$ &
$0.0\%$ \\ 
\cutinhead{Inverse power law potential (B), $\ome[0]{\phi} = 0.6$}
$8.4\times 10^{-8}$ & $-0.32$ & $-0.37$ &$12205$ & $12147$ & $0.5\%$ &
$267$ & $0.0\%$ \\* 
$9.9\times 10^{-2}$ & $-0.16$ & $-0.13$ & $10774$ & $10798$ & $0.2\%$
& $253$ & $0.2\%$ \\*
$0.22$ & $-0.1$ & $-8.1\times 10^{-2}$ & $10241$ & $10273$ & $0.3\%$
& $236$ & $0.2\%$ \\ 
\cutinhead{Pure exponential potential, $\ome[0]{\phi} = 0.6$}
$0.70$ & $ 0.00 $ & $ 7 \times 10^{-3} $ & $9014$ & $9049$ & $0.4\%$ &
$146$ & $2.3\%$  \\ 
\cutinhead{Pure exponential potential, $\ome[0]{\phi} = 0.2$}
 $0.22$ & $ 5\times 10^{-5} $ & $ 4.7 \times 10^{-3} $ & $9107$ & $9120$ & $0.1\%$ & $191$ & $0.3\%$  \\  
\cutinhead{Cosmological constant  (C), \ome[0]{\phi} = 0.6}
$0 $ & $ -1 $ & $ -1 $ & $13330$ & $13325$ & $0.0\%$ & $267$  &
$0.0\%$  \\ 
\cutinhead{Cold Dark Matter - no dark energy, \ome[0]{\phi} = 0}
$0$ & \nodata  &  \nodata   & $9133$ & $9133$ & $0.0\%$ & $201$  & $0.5\%$  \\ 
\enddata
\tablenotetext{a}{We compare analytical estimates with numerical solutions. We also list the
equation of state today $w_0$ (relevant for supernovae observations) and compare it with
the averaged equation of state \weff.
The cosmological parameters used can be read off \mbox{Table \ref{symb} }.}
\end{deluxetable}

\clearpage
\begin{deluxetable}{ccc}
\tablewidth{0pc} 
\tablecolumns{3}
\tablecaption{Symbols, their meanings and numerical values used in this paper.\label{symb}}
\tablehead{
\colhead{Symbol} & \colhead{Meaning} & \colhead{Value}  }

\startdata
$\Mp$ & reduced Planck mass $\Mp^{-2} \equiv 8\pi G$ & \nodata \\
$\scale$ & scale factor, normalised to unity today & \nodata \\
$a_{\rm ls}$ & scale factor at last scattering & $1100^{-1}$ \\
$h_0$ & Hubble parameter today $H_0 = 100\,h_0\,\textrm{km  s}^{-1} {\rm Mpc}^{-1}$ &  $0.65$ \\
\ome[0]{\rm r} & relativistic $\Omega$ today & $9.89\times 10^{-5}$ \\
\ome[0]{\rm b} & baryon $\Omega$ today & $0.05$ \\
$\bar c_s$ & $\tau$-averaged sound speed until last scattering & $ 0.52$ \\ 
$n$ & spectral index of initial perturbations & $1$ \\
\enddata
\end{deluxetable}

\end{document}